\documentclass[conference]{IEEEtran}

\usepackage{amsfonts}
\usepackage{amssymb}
\usepackage{stfloats}
\usepackage{cite}
\usepackage{graphicx}
\usepackage{epstopdf}
\usepackage{psfrag}
\usepackage{subfigure}
\usepackage{amsmath}
\usepackage{array}
\usepackage{stfloats}
\usepackage{multirow}
\usepackage{booktabs}
\usepackage{url}
\usepackage{graphicx}

\title{A Prototype Performance Analysis for V2V Communications using USRP-based Software Defined Radio Platform}
\author{{\em (Invited Paper)}\\
Fei Peng, Shunqing Zhang,~\IEEEmembership{Senior Member, IEEE}Shan Cao, 
and Shugong Xu,~\IEEEmembership{Fellow, IEEE}\\
\IEEEauthorblockA{Shanghai Institute for Advanced Communication and Data Science\\
Key laboratory of Specialty Fiber Optics and Optical Access Networks\\
Joint International Research Laboratory of Specialty Fiber Optics and Advanced Communication\\
Shanghai University, Shanghai 200444, China\\
E-mail: \{pfly\_shmily, shunqing, cshan, shugong\}@shu.edu.cn}
}

\begin{document}
\maketitle

\begin{abstract}
Autonomous driving is usually recognized as a promising technology to replace human drivers in the near future. To guarantee the safety performance in the daily life scenario, multiple-car intelligence with high quality inter-vehicle communication capability is necessary in general. In this paper, to figure out the potential practical issues in the vehicle-to-vehicle transmission, we present a software defined radio platform for V2V communication using universal software radio peripheral (USRP). Based on the LTE framework, we modify the frame structure, the signal processing mechanisms and the resource allocation schemes to emulate the updated LTE-V standard and generate the corresponding numerical results based on the real measured signals. As shown through some empirical studies, one to four dB back-off is in general required to guarantee the reliability performance for V2V communication environments.
\end{abstract}


\section{Introduction} \label{sect:intro}
Autonomous driving \cite{autonomous_driving}, as a promising approach to replace human drivers, has attracted significant research attention from industry and academy recently, especially after the explosive development of artificial intelligence. With the assistance from multiple sensors (camera, laser or millimeter wave radar),  single-car intelligence provides acceptable performance in the safety aspect. However, as reported in \cite{li2017analytical} and \cite{Gao2018}, the most challenging issue for autonomous driving is to provide guaranteed safety performance under all scenarios, where multiple-car intelligence or intelligent connected vehicles becomes critical and the high quality inter-vehicle communication will be necessary. 

In order to support fast and reliable inter-vehicle communication, traditional network oriented solution, such as long-term evolution (LTE), incurs notable transmission overhead between base stations and vehicle terminals, where the direct transmission among different vehicles will be a key enabler. Although the traditional {\em dedicated short range communications} (DSRC) solutions provides vehicle-to-vehicle (V2V) communication capability, the quality-of-service cannot be guaranteed, which triggers the new development of cellular-based vehicle-to-everything (V2X) solution such as LTE-vehicle (LTE-V) \cite{DSRC_vs_LTE-V1}. In the current literature, various kinds of resource allocation schemes have been proposed to improve the reliability and delay performance of LTE-V systems \cite{resource_allocation2}. For instance, in \cite{spectrum_sharing}, \cite{Multi-RAT_link_assignment} and \cite{Power_Allocation}, the wireless link level performance improvement using spectrum sharing, adaptive link assignment and power adjustment are reported respectively. In addition, the path loss, shadow fading and delay spread of the V2V channels have been analyzed in \cite{GuanIoTV2V}. Although the above results provide a comprehensive study in the numerical model based optimization framework, the following practical issues have {\em NOT} yet been investigated based on our present state of knowledge.

\begin{itemize}
    \item{\em Real-time Throughput and Reliability Evaluation} In the traditional model based evaluation method, the signal processing delay and reliability is usually ``assumed'' to be a constant number, and the error rate performance is a fixed value under given signal-to-noise (SNR) condition in general. However, in the practical vehicular communication scenario, the mobility induced channel variations and the processing complexity definitely impact the real-time throughput and reliability performance, which is critical for high reliable V2V transmission requirement.
    \item{\em Hardware Imperfectness} Another important factor to improve the V2V communication reliability is the pre-estimation of potential hardware impairments. Due to the high mobility of vehicles, the Doppler spread in the wireless environment incurs the highly dynamic receiving signals. As a result, the synchronization, the frequency offsets and the hardware mismatch issues become severe, which may impact the overall V2V transmission performance.   
\end{itemize}

In this paper, we present a software defined radio (SDR) platform for V2V communication using universal software radio peripheral (USRP). Based on the LabVIEW LTE framework develop by National Instrument (NI) \cite{NI_LABVIEW}, we modify the frame structure, the signal processing mechanisms and the resource allocation schemes to emulate the updated LTE-V standard and generate the corresponding numerical results based on the real measured signals. As shown through some empirical studies, we need a 1 to 4 dB transmit power back-off in order to guarantee the reliable block error rate (BLER) performance for V2V direct communication.

The rest of this paper is organized as follows. Section~\ref{sect:bg} provides a preliminary introduction of V2V communication and the associated standard progress. We demonstrate a software defined radio platform which supports V2V communication in Section~\ref{sect:ps} and provides the corresponding emulation results in Section~\ref{sect:sr}. Conclusions are given in Section~\ref{sect:conc}.

\section{Preliminary for LTE-V Systems} \label{sect:bg}

In this section, we provide a historical view of LTE-V related standard. In order to support device-to-device (D2D) communication in the cellular framework, the third generation partnership project (3GPP) proposes to introduce a new physical link, called {\em sidelink} in release 12, which allows two user equipments (UEs) to communicate directly without the support from base stations. However, to enable the direct V2V communication based on the sidelink concept, there are still many practical issues in the vehicular network environment.

To address the above practical issues, the distributed resource allocation mechanisms are first introduced in 3GPP release 14, where the frame structure is modified to support sub-channelization of the previous sidelink implementation and the energy sensing scheme is utilized before the direct V2V transmissions. In addition, the reference signal assignment is modified according to support the potential high mobility from V2V communication. On top of that, LTE-V use sensing based semi-persistent scheduling in the higher layer to reduce potential collisions.

\section{SDR Platform for V2V Transmission} \label{sect:ps}

In this section, we provide an overview of current platform and the corresponding modifications for V2V transmission. In addition, we also describe some enhanced features for V2V performance evaluation and further improvement.

\subsection{Platform Overview}

\begin{figure}[htbp]
\centerline{\includegraphics[width = 3.3in ]{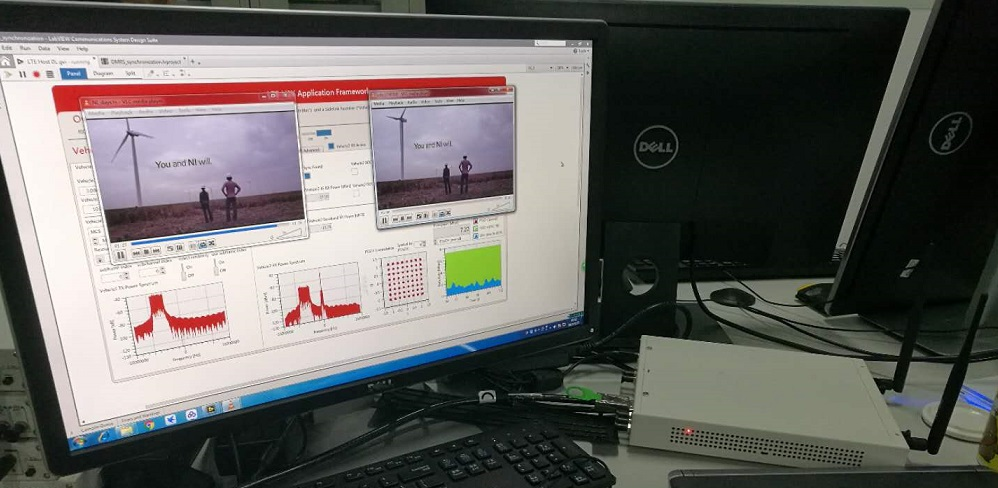}}
\caption{An overview of the software defined radio platform. In this platform, an FPGA-based programmable baseband processing unit and USRP-based radio frequency part is configured in the prototype system, where the video transmission applications are running through the air interfaces.}
\label{video_transmission_Demo}
\end{figure}

From the hardware point of view, the SDR platform contains a programmable baseband processing unit and a reconfigurable radio frequency peripheral as shown in Fig.~\ref{video_transmission_Demo}. To be more specific, the LabVIEW LTE framework is realized through a field programmable gate array (FPGA) based programmable baseband processing unit and a universal software radio peripheral (USRP) based radio frequency part, which covers the operating frequencies from 1.2 GHz to 6 GHz. In the software configuration, the initial version supported by LabVIEW LTE framework is 3GPP release 10, and the sidelink channel to support direct D2D/V2V transmission is required. 

\subsection{Adaptation to V2V Transmission}

As summarized in Section~\ref{sect:bg}, to support V2V direct transmission on top of 3GPP release 10 framework, we need to modify several processing parts as listed below.

\subsubsection{Frame Structure and Reference Signal}

\begin{figure}[htbp]
\centerline{\includegraphics[width = 3.5 in]{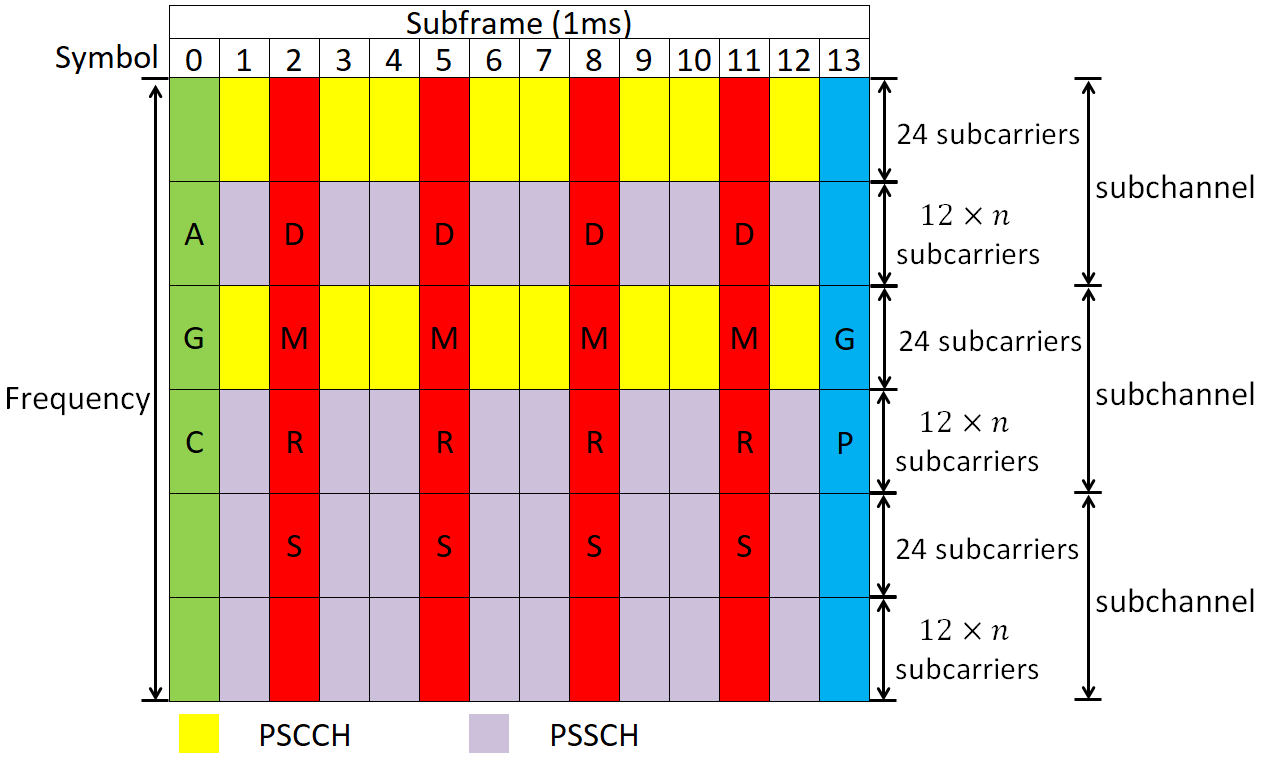}}
\caption{An illustration example of 3GPP release 14 frame structure for V2V Transmission, where the first and last symbols are used for automatic gain control and guard period respectively. In each sub-frame, four columns of DMRSs are inserted for channel estimation purpose and the remaining parts are used for V2V transmission. According to the size of data, one or several consecutive sub-channels can be selected to transmit data.}
\label{frame_structure_v3}
\end{figure}

To facilitate direct V2V transmission, the sidelink frame structure is modified to jointly process the control and data information in some specific symbols of each sub-frame according to 3GPP release 14 specification \cite{3gpp:36.213}, where the traditional approach to detect control information before data processing is no longer used. On the contrary, the whole sub-frame will be buffered and the corresponding control and data information is proceeded thereafter as shown in Fig.~\ref{frame_structure_v3}. Through this approach, each vehicle terminal can check the transmission status of each sub-frame before transmitting and the distributed resource allocation can be realized.  

Another important modification for V2V communication is the reference signal assignment. In order to reduce the impact of Doppler spread for high mobility vehicles, 3GPP release 14 \cite{3gpp:36.213} has specified to use up to four demodulation reference signals (DMRSs)\footnote{To be more specific, DMRSs are assigned in the $3^{\textrm{rd}}$, $6^{\textrm{th}}$, $9^{\textrm{th}}$ and $12^{\textrm{th}}$ symbols of each sub-frame respectively.} per sub-frame. In addition, LTE-V systems also reserve the first symbol for automatic gain control saturation and the last symbol for guard period. Note that the resource allocation in V2V transmission is selected on a sub-channel basis, where 48 subcarriers\footnote{For illustration purpose, we choose $n$ to be 2, and the extension to other standard values of $n$ is straight forward. We refer readers to \cite{3gpp:36.331} for more information.} are grouped as a basic configuration of each sub-channel in the frequency domain.

\subsubsection{Information Processing and Resource Allocation}

In the V2V transmission as explained before, the information processing and resource allocation is utilized on the sub-channel basis rather than the traditional resource block basis. In this sense, the calculation of resource elements for control and data information per sub-frame is quite different from the conventional LTE transmission. Meanwhile, since different number of consecutive sub-channels can be combined together, the control information extraction needs to be done per sub-channel basis and the data information can only be processed after the number of consecutive sub-channels has been finalized. As a result, the signal processing can be quite different from the conventional LTE solution.  

\begin{figure}[htbp]
\centerline{\includegraphics[scale=0.26]{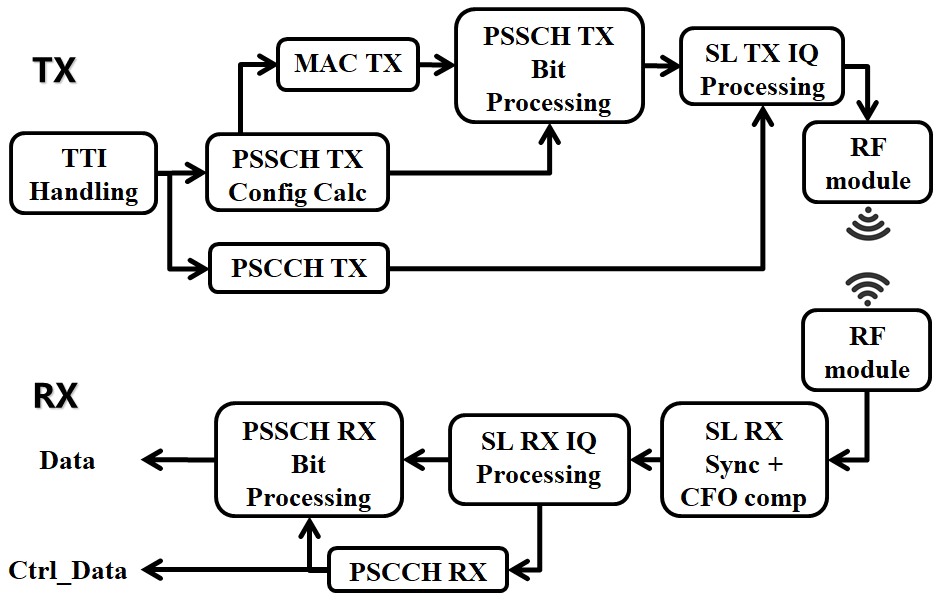}}
\caption{Block diagram of baseband processing in the SDR platform. The modification of frame structure and reference signal is mainly related to the module of SL IQ Processing in both transmitting and receiving part. The modification of information processing and resource allocation is mainly related to the module of PSCCH TX and PSCCH RX.}
\label{processing}
\end{figure}

In the prototype system as shown in Fig.~\ref{processing}, we modify the resource allocation part to support consecutive sub-channel assignment, where we allow random or pre-configured sub-channel assignment in each selection period\footnote{In this paper, we choose the selection period to be 1ms or 10ms and the extension to other standard compatible periods, such as 20ms, 50ms and 100ms, will be straight forward.}. In addition, we only allow adjacent mode for resource assignment, e.g. the physical sidelink control channel (PSCCH) and the physical sidelink shared channel (PSSCH) are transmitted subsequently. 

\subsection{Advanced Features}

To provide more insightful results and visualize the V2V transmission performance, we also introduce several advanced features in what follows.

\subsubsection{Performance Evaluator}
The first advanced feature that incorporated in this prototype system is the performance evaluator, which provides the possibility to track the system performance under different scenarios. Although to track the performance of real-time prototype systems is quite challenging in general, we design a fast caching scheme to temporally store the decoding results of V2V transmission and calculate the error rate performance in an offline manner. In this way, we are able to generate the reliability and throughput performance accordingly.

\subsubsection{Connection with NS-3}
Another important feature for the prototype system is the interaction capability with higher layer simulators. To make the V2V network evaluation feasible, we utilize the application programming interface on the LabVIEW platform to receive the upper layer traffic conducted by the network simulator version 3 (NS-3) \cite{ns3.org}. Through this approach, we are able to combine the LabVIEW physical layer with the upper layer communication process from NS-3, which allows a more reliable system level solution design and performance validation.

\section{Emulation Results} \label{sect:sr}

In this section, we provide some empirical results based on this SDR based V2V communication prototype system. To understand the imperfectness from wireless transmission, we build the indoor environment as shown in Fig.~\ref{video_transmission_Demo}, where the V2V communication is modeled through two separated antennas. Based on the LTE-V standard, we calculate the BLER and throughput of PSSCH as performance metrics. Other emulation parameters are summarized in Table~\ref{Block_error_rate_table}. 

\begin{table} [h] 
\centering 
\caption{Emulation Settings for SDR based V2V Communication Prototype System}
\label{Block_error_rate_table}
\footnotesize
\begin{tabular}{c c}  
\toprule
\textbf{Parameter}&\textbf{Value}\\
\midrule
Frequency & 5.9 GHz \\
\midrule
Modulation and Coding Scheme &  Level 0,5,10,15 \\
\midrule  
Subchannel size &  288 subcarriers \\
\midrule 
TX Power & -20 to -6 dBmW \\
\midrule 
Inter-Antenna Distance & 15 cm \\
\midrule 
Fading Environment & Indoor \\
\bottomrule
\end{tabular}  
\end{table} 

\subsection{Reliability}

In this experiment, we choose BLER versus transmit power relation to be the reliability measure. Different from the traditional deterministic software simulation, we calculate BLER performance in each second (with fixed transmit power) and repeat this experiment by 4000 times in order to get the statistics (average and standard deviation values) of this relation.   

\begin{figure}[htbp]
\centerline{\includegraphics[scale=0.221]{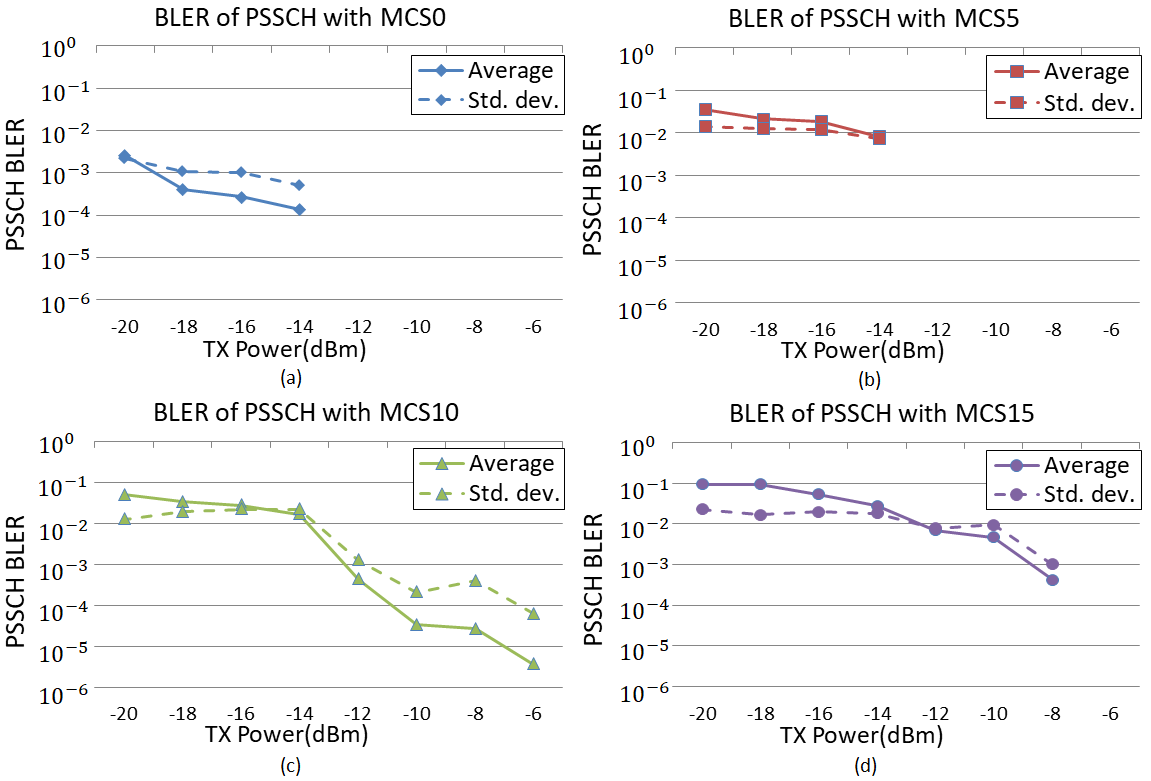}}
\caption{Average BLER and standard deviation versus transmit power relation for PSSCH transmission under different MCS levels. From this figure, we can observe that: the average BLER decrease monotonically with respect to the transmit power, while the standard deviation does not. By comparing the ratios between the standard deviation and average BLER under MCS level 10, we show that not only the average BLER value but also the statistics of BLER shall also be considered when the transmit power increases.}
\label{PSSCH_BLER_average_total}
\end{figure}

The average and standard deviation of BLER performance versus transmit power relation under different modulation and coding schemes (MCS) levels are shown in Fig.~\ref{PSSCH_BLER_average_total}. Intuitively, a higher transmit power results in smaller BLER value on average and a higher MCS level requires a higher transmit power to maintain the similar reliability performance in general. However, from the standard deviation point of view, we are {\em not} able to obtain the similar monotonical relation. If we further compare the ratios between the average value and standard deviation of BLER under MCS level 10, the overall results become 0.25 and 17.1 for the transmit power equal to -20 dBmW and -6 dBmW, respectively. In this sense, the average BLER may become unstable when the transmit power increases and in order to obtain a more reliable result, other statistics of BLER shall also be considered. 

\begin{figure}[htbp]
\centerline{\includegraphics[scale=0.17]{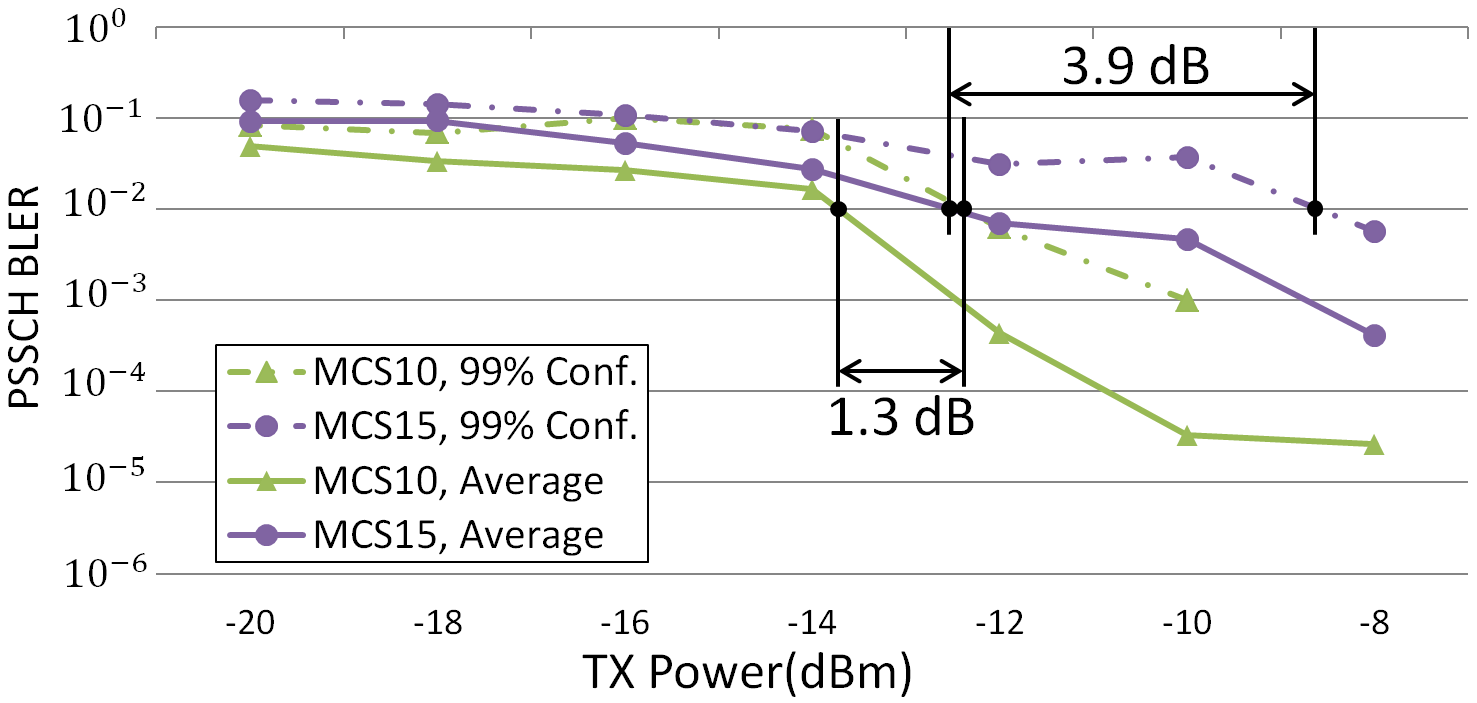}}
\caption{BELR versus transmit power relation for PSSCH transmission under different MCS levels.}
\label{The_BLER_at_0.99}
\end{figure}

In Fig~\ref{The_BLER_at_0.99}, we also compare the average BLER performance with the BLER value under 99\% confidence level. As we can see from this figure, a 1 to 4 dB SNR boosting will be necessary to ensure a reliable BLER result.

\subsection{Throughput}

\begin{figure}[htbp]
\centerline{\includegraphics[scale=0.196]{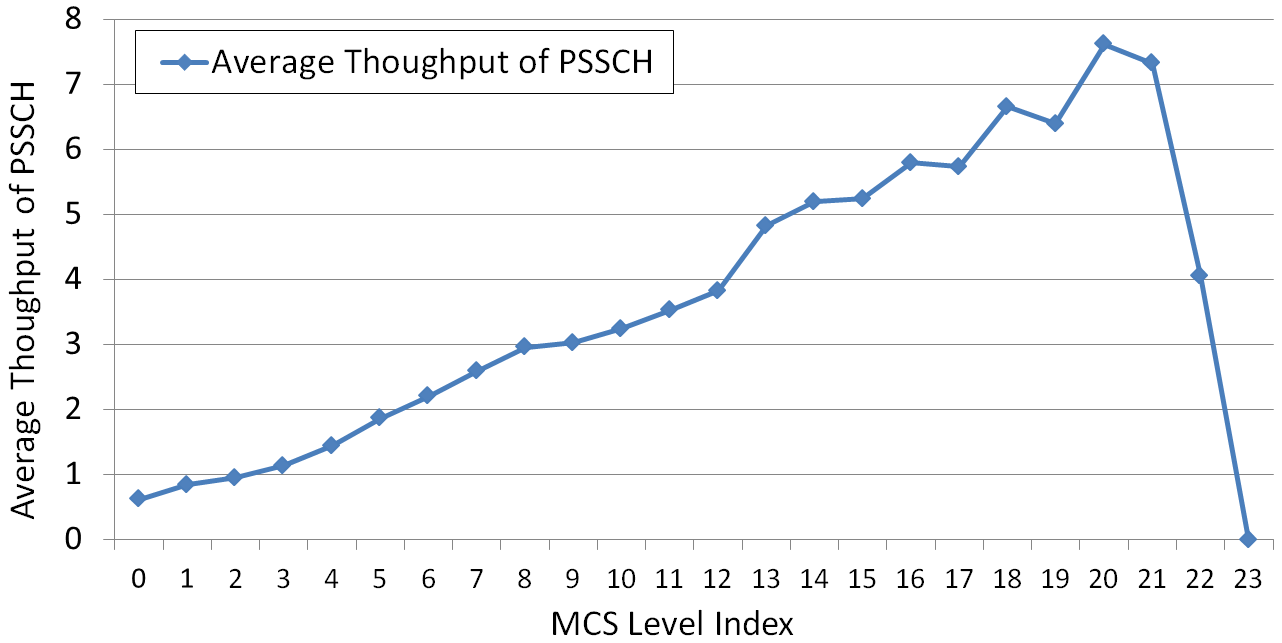}}
\caption{Average throughput of PSSCH versus different MCS level indices. The average throughput value of PSSCH channel increases monotonically in the lower MCS level region, while it drops rapidly when the MCS level is above 21. In the practical V2V communications, MCS adaptation would be more challenging, as it requires jointly consider the target BLER performance, the average throughput as well as the reliability requirement, where additional research efforts are worthwhile.}
\label{throughput}
\end{figure}

In the second experiment, we plot the average throughput versus MCS level relation to understand the achievable transmission rate of V2V communication. As shown in Fig.~\ref{throughput}, we are not able to obtain a monotonical relation between the average throughput and MCS levels when the packet drop happens in the high MCS levels. In other words, MCS adaptation is still necessary in the V2V communications. To be more specific, the non-safety messages with lower reliability requirement are more suitable for higher MCS levels, while the safety messages with higher reliability requirement need to trade off for lower MCS level. 

In the practical V2V communications, MCS adaptation would be more challenging, as it requires jointly consider the target BLER performance, the average throughput as well as the reliability requirement, where additional research efforts are worthwhile. 

\section{Conclusion} \label{sect:conc}

In this paper, we present an SDR based V2V communication platform using the LabVIEW LTE framework. In order to support direct V2V transmission, we modify the frame structure, the signal processing mechanisms and the resource allocation schemes to emulate LTE-V systems. According to the numerical results, we find that practical wireless environment and hardware imperfectness may result in a 1 to 4 dB performance loss in terms of BLER if a higher reliability needs to be guaranteed. In the future work, possible schemes will be designed to improve the error performance, and more practical scenarios will be considered for example making the experiment on moving vehicles. Meanwhile, we believe this platform will pave the way for future evaluation of V2V networks and 5G V2X communication technologies. 

\section*{Acknowledgement}
This work was supported by the National Natural Science Foundation of China (NSFC) Grants under No. 61701293, No. 61871262, the National Science and Technology Major Project Grants under No. 2018ZX03001009, the Huawei Innovation Research Program (HIRP), and research funds from Shanghai Institute for Advanced Communication and Data Science (SICS).

\bibliographystyle{IEEEtran}
\bibliography{IEEEfull,ref}

\end{document}